\documentclass[namedreferences]{kluwer}
\usepackage{epsfig}

\newcommand{\mic}{\mbox{$\mu$m}}
\newcommand{\ISO}{\mbox{\em{ISO}}}
\newcommand{\ISOSWS}{\mbox{\em{ISOSWS}}}
\newcommand{\ISOCAM}{\mbox{\em{ISOCAM}}}
\newcommand{\ISOPHOT}{\mbox{\em{ISOPHOT}}}
\newcommand{\ISOLWS}{\mbox{\em{ISOLWS}}}
\newcommand{\SCUBA}{\mbox{\em{SCUBA}}}
\newcommand{\IRAS}{\mbox{\em{IRAS}}}
\newcommand{\hii}{\mbox{H{\sc ii}}}
\newcommand{\hi}{\mbox{H{\sc i}}}
\newcommand{\hmol}{\mbox{H$_{\rm 2}$}}
\newcommand{\nature}{\mbox{{\it Nature\/}}}
\newcommand{\apj}{\mbox{{\it ApJ\/}}}
\newcommand{\aj}{\mbox{{\it AJ\/}}}

\newcommand{\apjl}{\mbox{{\it ApJLett\/}}}
\newcommand{\aap}{\mbox{{\it A\&A\/}}}
\newcommand{\araa}{\mbox{{\it ARA\&A\/}}}

\newcommand{\mnras}{\mbox{{\it MNRAS\/}}}
\newcommand{\cii}{\mbox{[C{\sc ii}]}}
\newcommand{\oi}{\mbox{[O{\sc i}]}}
\newcommand{\oiv}{\mbox{[O{\sc iv}]}}
\newcommand{\neii}{\mbox{[Ne{\sc ii}]}}

\newcommand{\nev}{\mbox{[Ne{\sc v}]}}
\newcommand{\isocol}{\mbox{[6.75$\mu$m]/[15$\mu$m]}}
\newcommand{\irascol}{\mbox{[60$\mu$m]/[100$\mu$m]}}

\begin{document}

\begin{frontmatter}

\end{frontmatter}

\begin{article}

\begin{opening}
    \title{A Mid and Far Infrared View of Galaxies}
    
    \author{C. J. \surname{Cesarsky}\email{ccesarsk@eso.org}}
    \institute{C.E.A. Saclay, DSM/DAPNIA/Service d'Astrophysique, 
    91191 Gif sur Yvette Cedex, FRANCE \\ European Southern 
    Observatory, Karl-Schwarzschildstr. 2, D-85754 Garching bei 
    M\"unchen, GERMANY}
    \author{M. \surname{Sauvage}\email{msauvage@cea.fr}}
    \institute{C.E.A. Saclay, DSM/DAPNIA/Service d'Astrophysique, 
    91191 Gif sur Yvette Cedex, FRANCE}
    
    \runningtitle{A MIR and FIR view of galaxies}
    
    \begin{abstract}
	From the disk of normal galaxies to the nucleus of prototype 
	active sources, we review the wealth of results and new 
	understanding provided by recent infrared probes 
	and, in particular, the four instruments on-board of \ISO.
    \end{abstract}
    \keywords{infrared: lines and bands -- infrared: continuum -- 
    galaxies: active -- galaxies: starburst -- stars: formation -- 
    ISM: infrared}
	
\end{opening}

\section{Introduction: infrared emission processes}
\label{sec:intro}

By allowing us a detailed view of the full 2-200 \mic\ 
spectral range, \ISO\ has dramatically increased our ability to 
investigate the processes giving rise to the spectral energy 
distribution (SED) of galaxies.  Yet before actually starting this 
review, let us use well-known Galactic sources to exemplify the links 
between SEDs and the physical state of the objects in which they 
occur.

\subsection{Mid-infrared bands and transient heating}
\label{subsec:transient}

The mid-infrared (MIR, 5-40 \mic) is distinct from the far-infrared 
(FIR) in the sense that it shows a large number of broad spectral 
bands (especially in the 5-15\,\mic\ range) and that most of the dust 
is out of thermal equilibrium, in a regime of transient heating.  The 
interesting property of transient heating is that, for a range of 
energy densities, the MIR flux scales linearly with the heating flux.  
This is clearly shown by \inlinecite{boulangerlisbon97} where they 
compare the MIR spectrum obtained on the peak of the NGC~7023 region 
(exposed to a B star radiation field) to that obtained on a diffuse 
cloud in the Chameleon region: the radiation fields differ by more 
than 3 orders of magnitude yet the MIR spectra are virtually 
identical.  More quantitatively, they have shown that this linear 
scaling remains valid for radiation fields lower than a few 10$^{3}$ 
times the solar neighborhood value.

Without expanding on the nature of the dust giving rise to this family 
of bands, it is worthwhile to list a few properties that \ISO\ has now 
firmly established.  Even at the high resolution of the \ISOSWS, the 
bands do not break up into a family of lines (as suggested e.g. by 
\opencite{leger89}).  In fact, in most regions of the ISM, the band 
profiles are very constant, and much better represented by Lorentzian 
than by Gaussian functions (see e.g. \opencite{boulorentz98}; 
\opencite{mattilangc89199}).  Given the broad wings of a Lorentzian, 
in regions where the bands are prominent, most of the detected flux 
actually comes from the band carriers.  As the debate still continues 
on the exact nature of the carriers, we will refer to them as the 
``infrared bands''.

Studies of Galactic regions also help to pinpoint the major 
sites of emission: although they are detected in diffuse cirrus clouds 
(\opencite{boulangerOph96}), most of the infrared band emission 
originates in the interface between \hii\ regions and molecular 
clouds, the photo-dissociation regions (PDR, see e.g. 
\opencite{cesarskym1796}; \opencite{verstraete96}).

\subsection{Continuum emission}
\label{subsec:continuum}

From 10-15~\mic\ to the submillimeter, the SEDs of most sources 
consist of a broad continuum.  This continuum cannot be fitted by a 
single back-body curve.  On the long wavelength side, this is due to 
the existence of more than one component of dust in thermal 
equilibrium.  On the short wavelength side, this is due to the 
transition from the transient heating regime to the thermal 
equilibrium regime: below a size threshold fixed by grain properties 
and the radiation field intensity, dust grains still undergo 
noticeable temperature fluctuations.  Depending on the heating 
radiation field, continuum emission will start to dominate over the 
infrared bands in the 12-15\,\mic\ range (e.g. in the PDR of M~17-SW, 
\opencite{cesarskym1796}), or even over the whole MIR range (e.g. in 
the \hii\ region of M~17-SW, \opencite{cesarskym1796}, see also the 
evolution of compact \hii\ region spectra in \opencite{coxunesco98}).  
One should note however that except in regions with a particularly 
hard radiation field, the MIR spectrum is generally dominated by the 
infrared bands.

\subsection{infrared lines}

The IR domain also gives access to very important diagnostic lines.  
These allow an almost extinction-free measurement of  
the intrinsic ionizing spectrum, the nature of the energy source, or 
the energetics of the interstellar medium (e.g. through the 
\cii\,158\,\mic\ or \oi\,63\,\mic\ lines).

The great advantage of ISO in this area is the possibility to observe 
the full set of lines from an object, free of any foreground emission, 
in a single fixed aperture (see e.g \opencite{colber99}, for M\,82).

\subsection{Active sources}

The three previous sections referred to emission processes occurring in 
the interstellar medium (ISM) of galaxies.  However, the IR is also a 
wavelength range where emission from active galactic nuclei (AGN) can 
be detected.  This can take two forms: (1) thermal emission from dust 
in the torus around the AGN, in which case we expect a very 
hot continuum as grains probably reach their sublimation temperatures, 
and (2) synchrotron emission from charged particles in the AGN strong 
magnetic field, in which case the emission takes the form of a 
featureless spectrum increasing with frequency.

\section{The interstellar medium in galactic disks}

A great deal of attention has been devoted to the characterization of 
nearby spirals' IR emission.  This has led to a deeper understanding 
of the distribution of dust in galaxies.

\subsection{Cold dust and the total amount of dust}
\label{subsec:cold}

Assuming that the \IRAS\ 60 and 100\,\mic\ fluxes from galaxies sample 
a single dust phase, dust masses and temperatures have been computed.  
Typical dust temperatures of $\sim$\,30\,K and dust masses one order of 
magnitude below that deduced for the Milky Way were obtained (e.g. 
\opencite{dy90}).  The advent of millimeter bolometers showed that 
this was likely an artifact of the wavelength domain sampled by \IRAS, 
but uncertainties in the exponent of the emissivity in the FIR/mm 
range prevented a definitive assessment of the question.

With the combination of \ISO\ and \SCUBA, this is now being resolved.  
The full FIR/mm SEDs of nearby spiral galaxies reveal a consistent 
picture: in normal galaxies (i.e. star-forming but not starbursting 
spirals) most of the emission longward of 150-200\,\mic\ is provided 
by a cold dust phase (T$\simeq$10-20\,K, \opencite{altonngc89198}; 
\opencite{bianchi98}; \opencite{daviesngc695699}; 
\opencite{haasM3198}; \opencite{israel891}; \opencite{kswc98}).  In 
fact, \inlinecite{altonngc89198} show clearly that the cold dust 
emission provides a significant fraction of the 100\,\mic\ flux.  This 
mixed nature of the 100\,\mic\ \IRAS\ band is very likely at the 
origin of numerous controversies on the interpretation of \IRAS\ 
luminosities.  Dust masses measurements now yield gas-to-dust 
ratios in external galaxies in the range 100-300, much more in 
agreement with the value of $\sim$160 measured in our galaxy.

These studies also point to a FIR/mm emissivity index of 2 (i.e. 
$\beta$ in $\kappa_{\nu}\propto\nu^{\beta}$) rather than the more 
commonly used 1.5\,.

\subsection{Spatial distribution of the IR emission}
\label{subsec:spatial}

Although all infrared maps show similar features, i.e. enhanced 
emission in the spiral arms and nucleus, a number of properties are 
emerging.  \inlinecite{alton98} and \inlinecite{daviesngc695699} have 
shown that \ISOPHOT\ 200\,\mic\ maps of normal spirals have a 
scale-length which is as large, or even larger than that of the stars.  
As the dust temperature probably decreases outwards, this implies that 
the scale-length of the cold dust column density is even larger.  
Comparison of radial profiles between cold dust traced at 850\,\mic\ 
with \SCUBA\ and the atomic and molecular components show that while 
the cold dust is more extended than the molecular gas, it is still 
less extended than the \hi\ gas (\opencite{daviesngc695699}).  
Furthermore the radial distributions of dust and molecular gas agree 
well in the central part of the disk (\opencite{israel891}).  This 
suggests that this cold dust is mostly associated with the molecular 
gas but that a non-negligible fraction also resides in the \hi\ gas.  
As we progress outward in the galaxy, the ISM becomes mostly atomic 
and the dust phase associated to \hi\ becomes more apparent.

Warmer dust, such as that detected by \IRAS\ or \ISOCAM, is clearly 
less extended than the cold phase and has a scale length smaller than 
that of the stars.  In NGC\,6946, \inlinecite{malhotra96} have shown 
that the \ISOCAM\ profiles at 6.75 and 15\,\mic\ have a scale length 
similar to that of the molecular gas and H$\alpha$ emission.  The 
60\,\mic\ scale length is intermediate between that of molecular gas 
and stars.  In fact it is close to that of the \hi+\hmol\ profiles, 
likely showing a transition between the warm and cold component.  In a 
thorough \ISOCAM\ study of the SW ring in M\,31, \inlinecite{pagani99} 
show a clear spatial correlation of the MIR emission with \hi\ and 
\hmol, while the correlation is much poorer with H$\alpha$.  This 
effect is also seen in NGC\,891 (\opencite{mattilangc89199}) and 
NGC\,7331 (\opencite{Smith98}).

The association of the MIR radiation with the \hi+\hmol\ gas in the 
inner part of the galaxies confirms that 
the emission originates mainly from the PDRs at the outer layers of 
interstellar clouds. Instead, the larger scale-length of the 
FIR/mm component points, for the colder dust, to an origin in the 
inner parts of interstellar clouds.

\subsection{Heating sources}
\label{subsec:sfr}

\begin{figure}
    \centerline{ 
    \epsfig{file=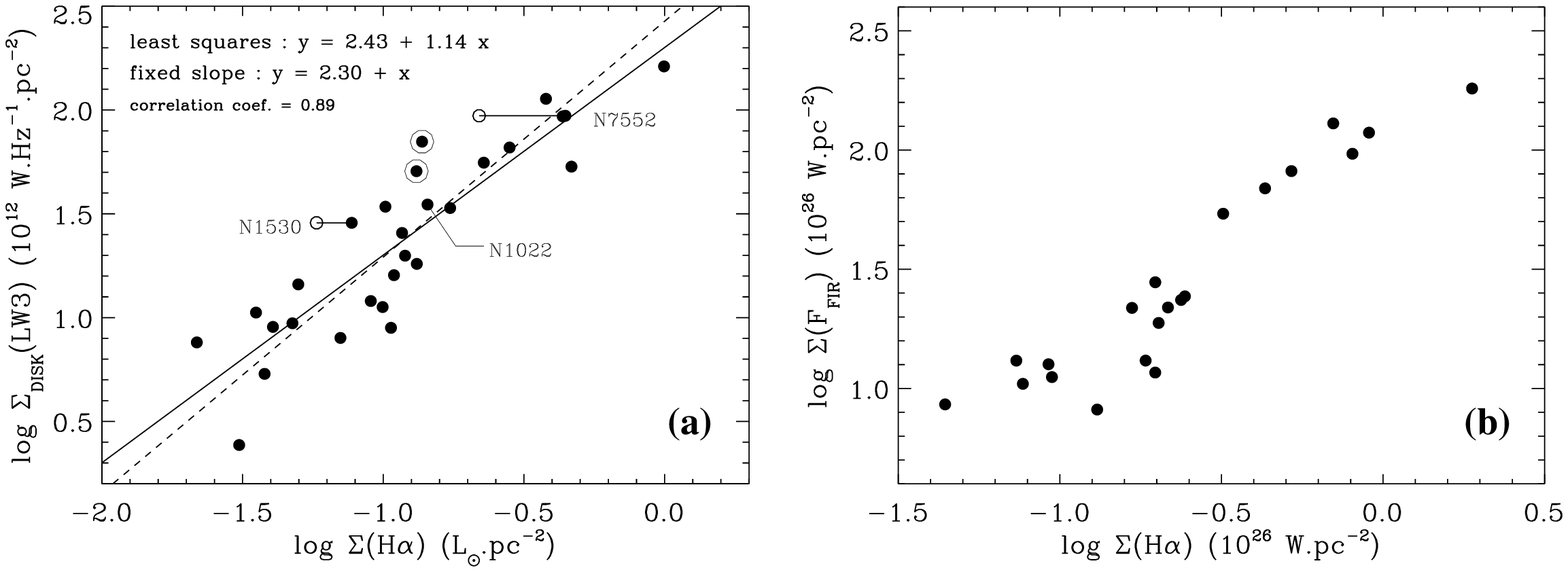,width=1.05\textwidth} 
    } 
    \caption{(a-left) The linear correlation between the 15\,\mic\ and 
    H$\alpha$ luminosities in the disk of spiral galaxies (normalized 
    by the disk's surface to cancel scale effects).  A similar linear 
    correlation is observed with 6.75\,\mic\ luminosities.  (b-right) 
    The correlation between the FIR and H$\alpha$ {\it total} fluxes 
    for galaxies that have little contribution of the central regions 
    in the 15\,\mic\ maps.  On this sample, the correlation is linear, 
    revealing a tight link between the two quantities (figures 
    adapted from Roussel et al., 1999b).}
    \label{fig:sfr}
\end{figure}

A long standing question has been the heating source for the IR 
emission, and in particular, the extent to which this emission can be 
considered a good star formation tracer (see e.g.  
\opencite{kennicutt_araa98}).  \ISO\ now allows us to re-address this 
question.  Examining the global luminosity of disks of spiral 
galaxies, \inlinecite{roussel99} establish a clear correlation between 
the MIR and H$\alpha$ luminosity of {\em disks} (see 
fig.~\ref{fig:sfr}a).  This correlation is linear and implies that in 
the disks of normal galaxies, the energy source for the MIR emission 
is the young stellar population and therefore that it can be used as a 
star formation tracer.  There are however a few caveats: (1) 
in the disk of normal spirals, the broad-band \isocol\ color is $\sim$ 
1, as expected if there is little if any contribution from very small 
hot grains such as observed in \hii\ regions (see 
sec.~\ref{subsec:continuum}).  This will not be the case in stronger 
star forming galaxies (see sec.~\ref{sec:diags}).  (2) The central 
regions of spiral galaxies typically have a lower \isocol\ ratio (e.g. 
\opencite{sauvagem5196}; \opencite{rousselunesco98}; 
\opencite{dale99}), possibly indicating again a higher local star 
formation rate.  Therefore, in the case of an unresolved galaxy, 
\inlinecite{roussel99} argue that their calibration should provide a 
lower limit to the actual star formation rate.

The possibility to separate the nuclear region from the disk of the 
galaxy in the MIR now allows a re-investigation of the FIR-H$\alpha$ 
correlation.  This correlation is known to be non-linear, implying 
that more than one process is present and that the FIR cannot be 
directly used to infer star-formation rates (e.g. \opencite{st92}).  
\inlinecite{roussel99} have selected in their sample those galaxies 
with the smallest contribution of the central region {\em in the MIR 
maps} and show that for this sample the FIR-H$\alpha$ correlation is 
very good and linear (fig.~\ref{fig:sfr}b).  This indicates that the 
non-linearity present in the general FIR-H$\alpha$ relation is induced 
by the nuclear component and that in the disk of normal galaxies, the 
FIR emission collected by \IRAS\ is mostly from dust heated by the 
young stellar population.

\section{Starbursts in the infrared}
\label{sec:starburst}

As already mentioned in section~\ref{subsec:sfr}, some galaxies have 
MIR colors that resemble those found in nearby \hii\ regions (e.g. 
compare \opencite{dale99}, to \opencite{contursi97}).  This indicates 
that the MIR-FIR properties of galaxies are useful tools to monitor 
star formation.

\subsection{Spectral energy distributions of starburst galaxies}
\label{subsec:specburst}

A first effect that can easily be understood is that the SED of a 
starburst galaxy is globally shifted toward higher dust temperatures.  
For instance, \inlinecite{kswc98} found that the starburst galaxies 
they studied did not require a cold (10\,K) dust phase in their SED. 
This is also confirmed by \inlinecite{klaasunesco98}, or in NGC 6090 
(\opencite{acostaunesco98}): the coldest dust phase required for these 
galaxies is 30-50\,K, in sharp contrast with what is obtained on 
normal spirals (see sec.~\ref{subsec:cold}).  One should however 
note that these temperature decompositions depend highly on the 
exponent of the dust emissivity assumed ($\beta$ = 1-2): 
\inlinecite{klaasunesco98} note that if $\beta$\,=\,2 were used for 
their sample, instead of $\beta$\,=\,1, there would be room for a 
colder dust phase.  Therefore, rather than attempting to determine 
precise temperature values, one should remember that in starburst 
galaxies, the peak of the SED shifts from 100-200\,\mic\ to the 
60-100\,\mic\ range.

The shift in the FIR is obviously reflected in the MIR. As the heating 
intensity rises, the small grain emission gradually shifts towards 
short wavelengths in the MIR window (see 
section~\ref{subsec:continuum}), producing a steeply rising continuum 
that can start anywhere in the 4-20\,\mic\ range.  It is important to 
realize that the wavelength at which the small grain continuum 
dominates over the infrared bands can vary and {\em is in fact most of the 
time beyond 12\,\mic}.  This can lead to some starburst galaxies being 
erroneously classified as normal from their \ISOPHOT-S spectrum (e.g 
\opencite{luunesco99}), which only extends to 12\,\mic.

Observations of known starbursts in the MIR also reveal that the 
infrared bands are rarely suppressed: all the star-forming regions of 
the Antennae show significant, if not dominant, emission from the 
infrared bands (\opencite{vigroux96}); the sample of galaxies with 
warm \IRAS\ colors selected by \inlinecite{mouri98} shows well-defined 
infrared bands but no sign yet of a rising continuum.  Finally, the 
template starburst spectrum used by \inlinecite{lsrmg98} (again from 
\ISOPHOT-S) is not markedly different from that of NGC~891 
(\opencite{mattilangc89199}).  Therefore, the clear signature of a 
starburst-powered galaxy in the MIR spectrum is more the presence of a 
strong continuum longward of $\sim$12\,\mic\ than the absence of 
infrared bands (see e.g. \opencite{crowther99} for a nice example on 
NGC\,5253)

Finally, an important point for starburst galaxies is the amount of 
extinction present on the line of sight.  Because infrared bands are 
located on both sides of the 9.7\,\mic\ silicate feature, a 
band-dominated extinction-free MIR spectrum will still show a 
depression around 10\,\mic.  Optical depth measurements are thus 
better made with line ratios (\opencite{lutz96}) and, in starburst 
galaxies, span a very wide range, $A_{V}\sim$10-100 
(\opencite{genzelULIRG98}).  More recently an attempt to measure the 
extinction by its imprint on the infrared band shapes has met with 
some success (\opencite{rigopoulouaj2000}) and confirmed the range 
mentioned above.

\subsection{Star formation in high IR luminosity galaxies}

A large amount of work has been devoted to the understanding of the 
physical processes at work in high IR luminosity objects, mostly 
interacting/merging systems.  Although, as mentioned in 
\inlinecite{sandersagn1999}, 3-4 of the 5 closest examples of 
ultraluminous IR galaxies (ULIRGs) contain a powerful if not dominant 
AGN, most ISO studies on luminous IR galaxies emphasize the importance 
of the starburst process in the generation of the IR luminosity (but 
see sec.~\ref{sec:diags}).  A very interesting point comes from 
studies of luminous and ultraluminous IR galaxies (the frontier being 
located at $L_{8-1000{\scriptscriptstyle {\mu}{\rm 
m}}}=10^{12}\,L_{\odot}$).

For their \ISOCAM\ sample of Luminous IR galaxies (LIG), 
\inlinecite{hwang98} show a clear anti-correlation between the 
compactness of the infrared source and the angular separation between 
the interacting objects.  This seems at odds with the absence of any 
correlation in the MIR properties of ULIRGs with the angular 
separation (but note that little IR imaging is available for ULIRGs).  
Similarly, \inlinecite{gaosol99} have shown that, in LIGs, there is a 
clear anti-correlation between the star formation efficiency (SFE) and 
the angular separation (i.e. higher SFE for closer pairs).  No such 
correlation is seen in the ULIRG sample of 
\inlinecite{rigopoulouaj2000}, but very interestingly, the maximum SFE 
reached by LIGs is of the order of the mean SFE of ULIRGs.  This seems 
to place ULIRGs as a limit-case for interaction triggered 
star-formation and may explain the lack of clear correlations with 
interaction parameters for the ULIRG sample.  It also supports the 
conclusions of e.g. \inlinecite{genzelULIRG98} and 
\inlinecite{rigopoulouaj2000} that the ULIRG phenomenon is mostly 
related to individual properties of the interacting galaxies and not 
directly to the interaction itself.

Of interest then are the spatially resolved observations of starburst 
galaxies and ULIRGs.  A common point of these studies is the discovery 
that in many cases, a significant part of the luminosity is produced 
by very compact, mostly extranuclear, sources.  This was first seen in the 
Antennae (\opencite{vigroux96}), but is now observed also in Mrk~171 
(\opencite{gallaisunesco98}), NGC~253 (\opencite{ketongc25399}), 
NGC~5253 (\opencite{crowther99}), or Arp~220 
(\opencite{soiferarp2201999}).  A plausible interpretation of these 
sources is that they are buried super-star-clusters, that will later 
evolve in the blue super-star clusters seen in interacting galaxies 
(e.g. \opencite{oconnel1994}).  Given the power output of these 
clusters, their infrared phase should be quite short, a fact that fits 
well in the scenario of starburst progression during the merging phase 
of interacting galaxies proposed by \inlinecite{rigopoulouaj2000}.

\subsection{The \cii\ deficit}
\label{subsec:ciidef}

One of the most surprising findings of \ISO\ comes from \ISOLWS: 
\inlinecite{malhotra97} observed that galaxies with the highest 
\irascol\ or star formation activity exhibited lower-than-expected 
\cii-to-FIR luminosity ratios.  This was unexpected given that \cii\ 
is predicted to be strong in regions exposed to far-UV photons that 
abonds in these galaxies.  This deficit was later confirmed by 
\inlinecite{luhman98} in a sample of ULIRGs.  Reasons for this deficit 
are still unknown.  Extinction or self-absorption have been rejected 
as the very large $A_{\rm V}$ required ($\sim$400-1000) are not 
confirmed by any other extinction measurements.  Favored explanations 
are (1) a decreased efficiency of photoelectric heating in very high 
UV fields (grains become positively charged or are destroyed, thus 
reducing the number of photo-electrons), or (2) softer than expected 
UV radiation fields in ULIRGs due either to a troncated initial mass 
function or the presence of aging starburst regions.

\section{Active galaxies}
\label{sec:agn}

Our understanding of active galaxies has advanced greatly in the recent 
years, mainly through multiwavelength observations of these objects.  
The infrared plays a particular role in these studies, for three
reasons:
 
\begin{itemize}
    
\item It is in the infrared that the spectra of these objects peak.  
Thus, a precise knowledge of the infrared flux is necessary to assess 
the bolometric luminosity, and the energetics of active galaxies.

\item Multiwavelength observations of AGNs tend to support 
a unified scheme, according to which these objects are intrinsically 
similar, with various scales, but are viewed under different angles.  
An energy source in their center is surrounded by a torus of obscuring 
gas and dust, with a radius of parsecs to tens or a hundred parsecs.  
The observational properties of the objects depend not only of the 
energy source, but also on the angle at which they are observed with 
respect of the torus.  In this scheme, radio quiet, steep spectrum 
quasars and Seyfert 1 galaxies are seen at intermediate angles, flat 
spectrum quasars and BL Lac are pole-on, and radio-galaxies and 
Seyfert 2 edge-on.  In the infrared, it is possible to observe the 
direct emission of the torus.  Models indicate that tori are still 
optically thick in the MIR (\opencite{pierkrolik1993}; 
\opencite{granato1997}).  But in the FIR a simple prediction of the 
unification models is that sources of similar energy should emit the 
same amount of thermal radiation from torus-heated dust.

\item The infrared is a domain so rich in spectral lines (see 
fig.~\ref{fig:circinus}), less subject to absorption than the 
optical ones, that infrared spectra permit indirectly to determine the 
ionizing continuum.

\end{itemize}

\subsection{Unification scheme and ISO observations}
\label{subsec:agnscheme}

Many \ISOPHOT\ results on quasars and radio-galaxies are still pending, 
but those available tend to support the unification scheme.  For radio 
quiet and steep spectrum QSOs, the SED has a bump at around 60\,\mic, 
and declines beyond 100\,\mic, as expected for multi-temperature dust 
emission in the range of tens to hundreds of K 
(\opencite{haasqso1998}).  Most interestingly, the prototype 
radiogalaxy Cygnus\,A exhibits a similar SED; but 3C20, with a similar 
radio flux, is not seen by \ISOPHOT. Also, on the whole, QSOs appear 
to be more luminous in the FIR than radio sources, thus, if the 
unification scheme is correct, the torus is also optically thick in 
the FIR. The SED of flat-spectrum QSOs is dominated by synchrotron 
emission, but for 3C279, a variable quasar, ``the (dust) bump pryes 
above the synchrotron spectrum" when the overall emission is low 
(\opencite{haasqso1998}).  In the case of Seyfert galaxies, the 
synchrotron emission is weak or absent in the IR. In the course of a 
study with \ISOPHOT\ of 10 CfA galaxies, \inlinecite{perezgarcia98} 
decompose the emission in two or three phases: warm (150\,K, 
corresponding to the nucleus), cold (40 to 50\,K, star forming 
regions), very cold (10 to 20\,K, cirrus).  As expected from the 
unification scheme, the warm phase is colder for Seyferts\,2 than for 
Seyferts\,1.  The respective extensions of the cold and the warm 
component in the IR are similar to those of the cold and the hot 
component in the R band.  But, in the FIR, the nuclear emission can be 
a substantial fraction of the total FIR, and thus of the total 
$L_{{\rm bol}}$ of the galaxy.

\inlinecite{clavel1998} did a statistical study with \ISOPHOT-S 
encompassing 26 Seyfert\,1 and 28 Seyfert\,2 galaxies.  They found 
that while Seyfert\,2 exhibit a weak continuum and strong infrared 
bands, Seyfert\,1 have 7 times stronger continua and weak or 
non-existing infrared bands.  This also agrees with expectations from 
MIR optically thick torus models.

\subsection{IR  lines in Seyfert galaxies and the Big Blue 
Bump}
\label{subsec:agnireuv}

\begin{figure}
    \centerline{
    \epsfig{file=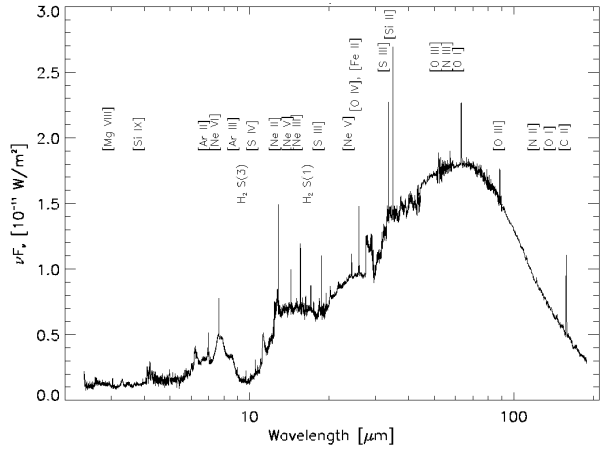,width=0.50\textwidth,angle=-90}}
    \caption{The combined \ISOSWS+\ISOLWS\ spectrum of the Circinus 
    galaxy (from Moorwood 1999).  Fine structure ionic lines as well 
    as some \hmol\ rotational lines are labeled.  Note also the 
    infrared bands in the 6-15\,\mic\ range as well as the probable 
    silicate absorption bands.}
    \label{fig:circinus}
    \end{figure}

With the spectrometers on board of \ISO, complemented by \ISOPHOT-S 
and the CVF of \ISOCAM, it has been possible to obtain a wide variety 
of results on Seyfert infrared spectral properties.  We display in 
fig.~\ref{fig:circinus} the combined \ISOSWS+\ISOLWS\ spectrum of 
Circinus, a prototype Seyfert 2 at 4\,Mpc 
(\opencite{moorwoodunesco98}).  There are 30 fine structure lines in 
this spectrum, emitted by species with ionization energies in the 8 to 
300\,eV range.  Note in particular the prominent high excitation 
lines of \oiv\ and \nev, mainly produced by the hard photons of 
AGNs.  Detailed modeling allowed \inlinecite{moorwoodcirc1996} to 
derive from this spectrum the ionizing EUV continuum, assuming that 
clouds are ionization bounded.  The result is a very hard ionizing 
continuum, with a very pronounced bump around 70\,eV. This Big Blue 
Bump is required by accretion disk models 
(\opencite{laor1990}), and can be considered the signature of a black 
hole.  However, a similar derivation made by 
\inlinecite{sturmngc41511999} on the Seyfert~1 NGC~4151, at 13\,Mpc, 
concluded that in the EUV spectrum of this galaxy there is a void, 
rather than a bump, at 70\,eV; they attribute this to absorption by 
neutral hydrogen placed between the narrow line region and the 
ionizing source.

\subsection{Spectral characteristics of active galaxies} 
\label{subsec:agnspec}

In addition to an active nucleus, Circinus is known to also have a 
circumnuclear starburst (\opencite{moorwoodcirc1996}); salient 
features of its spectrum (fig.~\ref{fig:circinus}), apart from the 
large number of fine structure lines, are the presence of infrared 
bands, the continuum bump due to hot dust and the broad silicate 
absorption around 9.7\,\mic.  In a subsequent study, using the 
\ISOCAM-CVF, \inlinecite{moorwoodunesco98} showed that the infrared 
band emission does not originate in the starburst ring, but is in 
fact caused by the remnants of an earlier starburst episode.

Another interesting case is that of Centaurus~A, the closest active 
galaxy, at only 3.1\,Mpc, which is suspected of hosting a misaligned 
BL~LAC nucleus (e.g. \opencite{baley1986}) at the center of an 
elliptical galaxy.  An \ISOCAM\ map by 
\inlinecite{Mirabelcena99} reveals the presence of a bisymmetric 
structure, similar to that of a barred spiral galaxy, in the central 
region of the giant elliptical galaxy.  This strange ``galaxy" is 
composed of the tidal debris of small galaxies accreted by the 
elliptical galaxy in the last $10^{9}$~years.  
\inlinecite{marconicena2000} recently observed this object with {\em 
NICMOS}, on board {\em HST}, and found Pa$\alpha$ emission indicative 
of enhanced star formation at the edges of the bar seen with \ISO, 
probably due to shocks associated with the bar.  The \ISOSWS\ spectrum 
of Centaurus~A (\opencite{alexanderunesco98}) is that of an AGN with a 
circumnuclear starburst, and has been modeled by these authors with a 
3.6\,pc torus inclined at 45$^{\circ}$.  \ISOCAM-CVF spectra from 5 to 
16\,\mic\ permit to distinguish very clearly between the nucleus 
dominated by a fast rising continuum, and the star formation regions 
emitting essentially in the infrared bands.

\section{Mid-infrared diagnostics to identify the nature of galaxies}
\label{sec:diags}

\begin{figure}

    \caption{(a-top left) The ISO-IRAS color diagram, that plots the 
    \isocol\ ratio versus the \irascol\ ratio.  Asterisks are for 
    starburst and active galaxies, diamonds for Virgo spirals, filled 
    circles for barred spirals and open squares for blue compact 
    galaxies.  (b-right) The Genzel et al.  (1998) diagram that 
    combines the \oiv/\neii\ ratio with the 7.7\,\mic\ 
    line-to-continuum ratio.  AGN-dominated galaxies have small 
    7.7\,\mic\ L/C ratios and large \oiv/\neii\ ratios.  Known 
    starburst are plotted as triangles, known AGNs as crossed squares 
    and their sample of ULIRGs as filled circles.  (c-bottom left) The 
    Laurent et al.  (2000) diagram for \ISOCAM\ data, that plots the 
    ratio of the 15\,\mic\ band to the 6.75\,\mic\ band versus the 
    ratio of the 6.75\,\mic\ band to the 6\,\mic\ band.  Galaxies from 
    the \ISOCAM\ central program are placed in that diagram.  Large 
    symbols represent objects with a known AGN that, expectedly, fall 
    in the AGN corner of the diagram.}

    \label{fig:diagplot}
    \end{figure}
    
The first such tool is one that combines \ISO\ and \IRAS\ information 
in the so-called \ISO-\IRAS\ color diagram.  It compares the 
\isocol\ ratio from \ISOCAM\ to the \irascol\ ratio from \IRAS\ (see 
fig.~\ref{fig:diagplot}a).  It is a first step in assessing the 
nature of \ISO\ galaxies.  For a large range of \irascol\ colors, the 
\isocol\ color is roughly constant.  This is the space occupied 
by normal star forming galaxies.  It is only beyond an \IRAS\ color 
$\geq$-0.2 that the \isocol\ color decreases.  Only blue compact, 
interacting or starburst galaxies occupy that part of the diagram, an 
expected fact from the previous discussion: the radiation field is high 
enough that the small grain continuum has been shifted into the 
\ISOCAM\ band.

This region of the \ISO-\IRAS\ color diagram is obviously of high 
interest as it hosts galaxies providing most of the IR energy 
collected in the Universe, and it has therefore been explored in more 
details.  \inlinecite{genzelULIRG98} were very succesful in arranging 
a sample of 13 ULIRGs on a plot representing the ratio \oiv/\neii\ 
versus the relative strength of the 7.7\,\mic\ infrared band 
(fig.~\ref{fig:diagplot}b).  In this diagram, the ULIRGs tend to lie 
close to the starburst region, but some clearly contain an energetic 
AGN. Similar analyzes, using only the 7.7\,\mic\ line-to-continuum 
(L/C) tool on larger samples were presented by \inlinecite{lsrmg98} 
and \inlinecite{rigopoulouaj2000} from which they conclude that indeed 
the fraction of ULIRGs powered by an AGN increases with the infrared 
luminosity, but also that ULIRGs are predominantly 
($\sim$70-80\%) starburst powered.

This 7.7\,\mic\ L/C tool is however ambiguous as some starbursts have 
no infrared bands while some AGN exhibit strong bands.  Thus the 
\inlinecite{laurentdiag2000} diagnostic (fig.~\ref{fig:diagplot}c), 
working on the broader \ISOCAM\ band, uses the flux ratio of the broad 
6.75\,\mic\ to the broad 15\,\mic\ band compared to the flux ratio of the broad 
6.75\,\mic\ band to the narrow 6\,\mic\ band to make a finer 
distinction between AGN, starburst and normal star-forming regions.  
Given that AGNs and starbursts have very different continuum shapes, 
this tool is very successful in distinguishing one from the other.  
Since AGNs have much more flux in the 6\,\mic\ range than starburst or 
star-forming galaxies, a ``band-less'' starburst is not mistaken for 
an AGN. Applying this tool to ULIRGs show that a fraction larger than 
that identified by e.g. \inlinecite{rigopoulouaj2000} is AGN-powered.  
This type of method has great potential for future studies with the 
NGST, if its wavelength range is sufficiently 
extended.

\section{Deep surveys}
\label{sec:deep}

Gains in sensitivity with \ISOCAM, compared to \IRAS\ at 12\,\mic, 
have made it possible to extend the range of MIR counts by three 
orders of magnitude.  Similarly \ISOPHOT\ allowed to perform, for the first 
time, surveys at 175\,\mic, a range of considerable cosmological 
interest.  The ISO counts give a radically new view of star formation 
in the universe between now and $z\sim$2.

\subsection{MIR templates and K~correction effects}
\label{subsec:deeptempl}

\ISOCAM\ surveys have essentially been performed in two filters, 
around 6.75 and 15\,\mic\ (but see also the 12\,\mic\ survey of 
\opencite{clements12mic99}).  Throughout this review, we have 
presented, discussed or alluded to template spectra of the various 
galaxy types in the MIR. For nearby objects, the 6.75\,\mic\ filter 
probes infrared bands, while the 15\,\mic\ filter probes 
preferentially warm dust and neon lines.  As the redshift increases, 
the infrared bands are more and more shifted to the 15\,\mic\ filter.  
For galaxies intrinsically bright in these bands, the K correction at 
15\,\mic\ is positive (sources appear to be fainter with increasing 
distances), but flat between $z\simeq0.4$ and 1.3\,.  For objects with 
$z>0.4$, the 6.75\,\mic\ filter has access only to the starlight or to 
an eventual AGN contributions; the same is true at 15\,\mic\ if 
$z>1.5$ (see \opencite{ausselhdf99}; \opencite{elbaz2000}).

The situation is even more favorable for surveys with ISOPHOT at 
175\,\mic; there, since most galaxy SEDs peak well below this 
wavelength, as $z$ increases, the K correction is negative, favoring 
detection of distant galaxies.

\subsection{The surveyed regions}
\label{subsec:deepregion}

The fields surveyed have been selected for their low zodiacal and 
cirrus emission; the second point is particularly important for the 
FIR studies.  They can be found, in the Northern 
hemisphere, in the Lockman Hole, and in the Southern hemisphere, in 
the Marano field.  Both fields have been the subject of 
multiwavelength studies, but often not to the depth required to 
interpret the results of \ISOCAM\ surveys.  These surveys are 
therefore well complemented by studies on well-known fields, the HDF 
North and a CFRS field at 6.75 and 15\,\mic, and the SSA13 field at 
6.75\,\mic.  Few results are available on the other fields at 
6.75\,\mic, where observations at other wavebands, not yet available 
for many of the surveys, are crucial to avoid contamination by 
galactic stars (but see \opencite{taniguchilock97}; 
\opencite{flores98}; \opencite{satounesco98}), and for lack of space 
we concentrate here, for ISOCAM, on 15\,\mic\ results.

\subsection{The nature of the ISOCAM 15\,\mic\ galaxies in a CFRS 
field and in HDF-North}
\label{subsec:deepnature}
 
Combining the deep survey on a CFRS field (\opencite{florescfrs15_99}) 
and that on the HDF North (\opencite{ausselhdf99}), 83 galaxies have 
been detected at 15\,\mic, with fluxes in excess of 250\,$\mu$Jy for 
the CFRS, and $>$100\,$\mu$Jy for the HDF. The positional accuracy of 
6$"$ allowed in almost all cases to identify an optical counterpart, 
brighter than I(Kron-Cousins)$\simeq$23; the mean redshift is 0.7.

In the CFRS field, Flores et al.  derived the type of the galaxies for 
which they had enough spectral information, including radio fluxes, by 
comparing their SED to local templates compiled by 
\inlinecite{schmitt1997}.  They found that more than two thirds of 
these galaxies are starburst or post-starburst, and that at least at 
$z\sim0.4$ to 0.6, their overall contribution to the global star 
formation rate is dominant.

Aussel et al., (1999a,b) compared the 27 MIR HDF sources with known 
redshift to a reference sample of HDF sources from an optical 
catalogue.  While the redshift distribution of the two samples are 
similar, the color distributions are very different.  \ISOCAM\ picks 
up galaxies with B-I colors corresponding to spirals, and is blind to 
the faint blue population responsible for the excess in the B counts 
(\opencite{ellis1997}).  In summary: it appears that most of the MIR 
galaxies have luminosities of the order or brighter than $L_{\rm 
bol}\simeq5\,10^{11}\,L_{\odot}$ (i.e. LIGs but not ULIGs), and are 
spiral or merging systems with normal colors, harboring an obscured 
starburst or a very well hidden active nucleus.  

\subsection{Source counts at 15\,\mic}
\label{subsec:deepcounts}

\begin{figure}
    \centerline{
    \epsfig{file=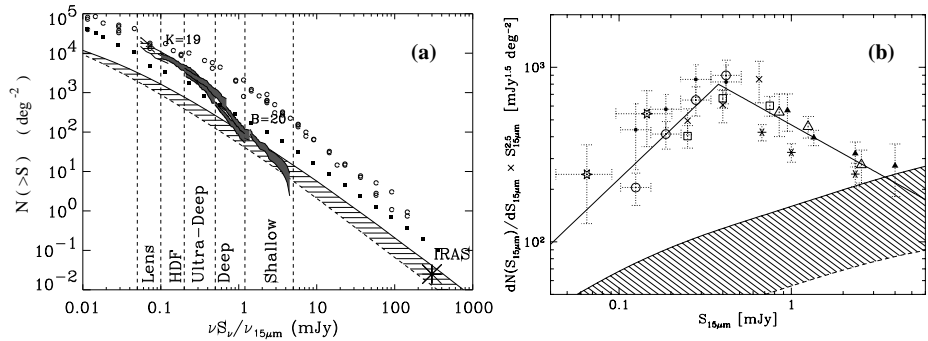,width=1.05\textwidth}
    }

    \caption{(a-left) A summary plot of the integral counts of 
    \ISOCAM\ 15\,\mic\ surveys with 68\% confidence contours.  K 
    counts (Gardner et al., 1993) and B counts (Metcalfe et al., 
    1995), multiplied by $\nu/\nu_{15\mic}$ to represent the relative 
    energy densities at high fluxes are overplotted with circles and 
    filled squares, respectively.  The hatched area materializes 
    expectations from models assuming no evolution and normalized to 
    the IRAS 12\,\mic\ local luminosity function (see Elbaz et al., 
    1999 for details).  (b-right) Differential number counts of 
    15\,\mic\ galaxies (also from Elbaz et al.  1999).  The counts 
    are normalized to an euclidian distribution of non-evolving 
    sources which would have a slope of $\alpha=-2.5$ in such a 
    universe.  The hatched area has the same meaning as in (a).}

    \label{fig:counts}
    \end{figure}

Thanks to the \ISOCAM\ surveys, it has been possible to extend by 
three orders of magnitude the $\log N-\log S$ diagram of MIR sources 
with respect to the IRAS results.  Fig.~\ref{fig:counts}, from 
\inlinecite{elbaz2000}, shows the integral and the differential counts 
derived from the Lockman Hole and the Marano shallow, deep and 
ultra-deep surveys performed on guaranteed time (IGTES, PI 
C.Cesarsky), the two HDF surveys, and an extension to fainter fluxes 
through the study of a lensed galaxy cluster 
\inlinecite{altierilens99}.  These counts are complemented, at flux 
$>$2\,mJy, by the results of ELAIS (PI M. Rowan Robinson; see 
\opencite{serjeant2000}).  Five independent surveys confirm the breaks 
of these curves at 400\,$\mu$Jy; in the differential plot, the slope 
passes from (-1.6) to (-3) for faint sources.  On the same figures are 
shown the expectations from models assuming no evolution, which take 
into account the infrared bands in the galaxy spectra.  Despite the 
uncertainties in these curves (see discussion in 
\opencite{elbaz2000b}), there is no question that at least at fluxes 
below 400\,$\mu$Jy, these source counts imply strong evolution in the 
population of MIR sources.

The \ISOCAM\ counts also indicate that the sources responsible for 
them are not ULIRGs with spectra like Arp\,220; otherwise, they would 
produce an infrared background at 140\,\mic\ in excess of the value 
observed by {\em COBE} (\opencite{puget1996}).

\subsection{Source counts at 175\,\mic}
\label{subsec:firback}

Deep counts with \ISOPHOT\ at 175\,\mic\ in the Lockman Hole and in 
the Marano field (\opencite{kawaraunesco1998}; \opencite{puget1999}), 
down to 120\,mJy, also give a $\log N-\log S$ curve with a steep slope 
and a source density which is higher by a factor of order 10 than 
extrapolations from \IRAS\ or predictions of no evolution models.  For 
these sources, identifications are uncertain, given the large error 
boxes.  Because of the large negative K~correction at this wavelength, 
it is likely that some of the sources seen are at $z$ in the range 1 
to 2.  Indeed, using the \ISOCAM\ results, one can show that they 
cannot be a population of galaxies with average redshift 0.7 and a 
spectrum like M\,82 or Arp\,220, since in that case \ISOCAM\ would see 
many more sources in the range $>$1\,mJy (\opencite{elbaz2000b}).  
Thus, while a fraction of the sources are in common, on the average 
\ISOCAM\ and \ISOPHOT\ sample different populations of sources.  While 
\ISOCAM\ pinpoints those of the B and K sources, at $z<1$, that 
harbor activity hidden by dust, \ISOPHOT\ misses many of them but can 
find more distant far-infrared beacons.

\medskip

A conclusion on such a broad subject is hard to write, when a large 
number of our colleagues are devoting most of their time to a finer 
analysis of the \ISO\ data.  Let us instead remark that the areas 
where new infrared data turn out to be of key importance are growing 
day by day, strongly supporting future infrared missions.

\begin{acknowledgements}
    We are grateful to the organizers for exerting the right pressure 
    resulting in the timely delivery of the manuscript, and to the 
    conference sponsors, the Anglo-American Chairman's fund and 
    SASSOL, for allowing this very interesting meeting to take place.
\end{acknowledgements}

\end{article}

\end{document}